\documentstyle[12pt]{article}
\begin{document}
\vspace*{-1in}
\begin{flushright}
SINP/TNP/94-13\\
CUPP-94/3\\
\end{flushright}
\vskip 70pt
\begin{center}
RESONANT NEUTRINO OSCILLATIONS \\
AND SHOCK REVIVAL IN TYPE-II SUPERNOVAE \\
\vskip 30pt
{\it Srubabati Goswami $^{a}$, Kamales Kar $^{b}$, Amitava
Raychaudhuri $^{a}$ \\
$^{a}$ Department of Pure Physics, University of Calcutta,\\
92 Acharya Prafulla Chandra Road, Calcutta 700009, INDIA.\\
$^{b}$ Saha Institute of Nuclear Physics,\\1/AF, Bidhannagar,
Calcutta 700064, INDIA.\\}

\vskip 60pt

{\bf ABSTRACT}
\end{center}

The role of matter enhanced resonant neutrino oscillations
in reviving a stalled shock
in a type-II supernova through delayed neutrino heating
is investigated.  The extent of neutrino heating
is estimated for the allowed possibility of complete flavour conversion
self-consistently with the changes in nuclear equilibrium.
The average internal energy per nucleon is substantially increased
indicating the possiblty of a robust explosion.

\vskip 25pt
\begin{center}
\end{center}
\vskip 40pt
\begin{flushleft}
November 1994\\
\end{flushleft}
\newpage
\setcounter{page} 1
\pagestyle{plain}

The theory of supernova explosions has seen a lot of progress in the
last fifteen years but it is still not able to deliver an explosion
with the right energy. Though most of the ingredients developed for
the scenario of gravitational collapse and shock propagation after core
rebounding are believed to be correct and some have got confirmation
through the detection of neutrinos of SN1987A, the strong dissipation
of shock energy through nuclear dissociation and neutrino losses make
the shock either fail to reach the edge of the core or emerge, after
neutrino heating, with energies less than 10$^{51}$ ergs, a
factor of 3 to 4 less than observed values. Different mechanisms like
improved pre-supernova conditions, soft equation of state, general
relativity at high densities, convection and improved neutrino
physics have all been invoked to solve this problem. In this letter we
describe a novel scenario by coupling neutrino flavour conversion in
matter (the MSW mechanism) with the delayed neutrino heating that can
deliver extra energy to matter behind the shock to generate a more
robust explosion. Fuller {\it et. al.} \cite{fuller} noted this possibility
of conversion of mu/tau neutrinos to electron type between the
neutrino-sphere and the stalled shock for some ranges of neutrino mixing
parameters and estimated the extra heating and the explosion energy.
We incorporate this heating to a semianalytic evolutionary calculation
of the thermodynamic conditions behind the shock front of a $25M_{\odot}$
star and demonstrate explicitly the change in  nuclear statistical
equilibrium (NSE). We also calculate the gain in internal energy achieved
by the increased heating induced by neutrino flavour oscillations.
The results are encouraging.

At present there are two possible scenarios for supernova explosion.
For stars in the mass range $8M_{\odot} \leq M \leq 15M_{\odot}$,
under some very special conditions on the size and structure of the
core and the equation of state, the shock continues its outward
propagation and the star explodes within some tens of milliseconds
after the beginning of collapse. This is the prompt explosion scenario
\cite{prompt}. For more masssive stars the energy of the shock gets
dissipated in dissociating nuclei and producing $\nu \bar{\nu}$ pairs
and the shock stalls at a radius of a few hundred kms and becomes an
accretion shock. It is subsequently revived by the heating caused by
neutrinos from the neutrino-sphere. This is the delayed explosion
mechanism \cite{wilson,bw85}. This late time neutrino heating behind
the shock is caused by their absorption reactions on the nuclei as well
as free nucleons and by charged and neutral current scattering reactions.
Ray and Kar \cite{rk87} considered a schematic model for
delayed neutrino heating with matter in the form of a typical heavy
nucleus $^{56}{Fe}$, $\alpha$-particles, neutrons and protons.
They  calculated the amount of neutrino heating treating changes in the
NSE behind the shock front in a self-consistent fashion.
Subsequently, Haxton \cite{haxton} pointed out the
necessity of including the neutral current inelastic scattering of nuclei
to giant resonance states in calculating the neutrino heating rates.
Although all three flavours participate in this, the $\nu_{\mu}$
and $\nu_{\tau}$ neutrinos contribute more to the heating by the above
process because of their higher equilibrium temperature. In \cite{haxton},
the charged current cross-sections of $\nu_{e}$s and $\bar{\nu_e}$s on
nuclei for some typical neutrino temperatures are also tabulated.
Ray and Kar \cite{rk92} incorporated these cross-sections in their
semianalytic model \cite{rk87} and computed the thermodynamic state of
of matter behind the shock as a function of time and estimated the heating
rates. Fuller {\it{et. al.}} \cite{fuller}
pointed out that matter enhanced oscillations of neutrinos in
the region between the neutrino-sphere and the stalled shock can
increase the heating rate appreciably and may result in a
delayed explosion with an energy $\geq 10^{51}$ ergs, for a
cosmologically significant $\nu_{\mu}$ or $\nu_{\tau}$ mass of 10
-100 eV and small vacuum mixing angles.
Due to neutrino flavour mixing $\nu_{\mu}$s or $\nu_{\tau}$s
(which carry a higher average energy)
get converted to $\nu_e$s, the upshot being the production of higher
energy  $\nu_e$s which can heat the shock more effectively.

We re-examine the situation modelled in \cite{rk92} with the addition
of neutrino heating in the presence of matter induced neutrino oscillations.
We consider a thin layer of matter behind the shock, at a radius $R_{m}$,
which gets heated by the neutrinos coming from the neutrino-sphere
situated at a distance $R_{\nu}$ from the center.
Matter enhanced resonant neutrino oscillations
have been considered extensively in the context of the solar neutrino
problem \cite{kuo}. The intrinsic neutrino properties needed for the
purpose are neutrino mass squared difference, $\Delta{m^{2}}$, and the
mixing angle in vacuum, $\theta_{V}$. Oscillation of neutrinos in matter
is different from that in vacuum because interactions
modify their dispersion relation and consequently
they develop an effective mass dependent on the matter density.
The mixing angle in matter can be expressed as
\begin{equation}
\tan{2\theta_{M}}~=~{\tan2\theta_{V}}/{(1-E_{\nu}/E_{A})}
\label{tan2thetam}
\end{equation}
$E_{\nu}$ is the neutrino energy; $E_{A}$ is given by
$E_{A} = {\Delta{m^{2}}\cos2\theta_{V}}/{2\sqrt{2}G_{F}n_{e}}$,
where $n_e$ denotes the electron density.
As is seen from (\ref{tan2thetam}) $\tan{2\theta_M}$ goes through a
resonance when $E_{\nu} = E_{A}$. From this
one gets an expression for the resonance density as \cite{fuller},
\begin{equation}
\rho_{res} = (2.108 \times 10^{10} {\rm gm/cc})\left(\frac{0.5}{Y_e}\right)
\left(\frac{\Delta{m^2} \cos{2\theta_{V}}}{1600 {\rm eV^{2}}}\right)
\left(\frac{1 {\rm MeV}}{E_{\nu}}\right)
\label{resden}
\end{equation}
At resonance, $\theta_{M}$ is $\pi/4$ and one gets maximal flavour mixing.
The effectiveness of neutrino oscillations in increasing the heating rate
depends crucially on whether a resonance is encountered between the
neutrino-sphere and the shock front or not.
If a neutrino of flavour $f$ encounters a resonance between the
neutrino-sphere and the shock then its probabiity to remain a
$\nu_{f}$ at the position of the shock is given as,
\begin{equation}
P(\nu_f \rightarrow \nu_f) = 0.5 + (0.5 -
P_J)\cos{2\theta_{R_\nu}}\cos{2\theta_{R_{m}}}
\label{pnuf}
\end{equation}
where $\theta_{R_\nu}$ ($\theta_{R_{m}}$) is the neutrino mixing angle
in matter at the position of the neutrino-sphere (shock).
$P_{J}$ is the non-adiabatic transition probability between the
two neutrino states and can be expressed in the Landau-Zener
approximation as,
P$_{J} = exp(-{E_{NA}}/{E_{\nu}})$ where,
\begin{equation}
E_{NA} = (0.026 {\rm MeV}) \frac{{\sin^2}2{\theta_{V}}}{\cos2\theta_{V}}
\left(\frac{\Delta{m^2}}{1 {\rm eV^{2}}}\right)
\left(\frac{r_{res}}{1 {\rm cm}}\right)
\label{ena}
\end{equation}
$r_{res}$ is the resonance radius. $E_{NA}$ can be obtained using
(\ref{resden}) and a density profile \cite{bethe}
$\rho$=${(10^{31}{\rm gm/cc})}(r/{\rm 1 cm})^{-3}$.
The above equations are for oscillation
between two neutrino flavours. In what follows we take these two
flavours as the electron and the tau neutrino.
A neutrino passing through the resonance density can
undergo complete flavour transformation for suitable values of the
parameters. A total conversion of the $\nu_{\tau}$, which carry a
higher energy, to $\nu_{e}$
between the neutrino-sphere and the shock will generate maximum
heating. Ideally, one should couple the neutrino
flavour transformations with the hydrodynamics of the shock and
calculate the probabilities of conversion between the neutrino-sphere
and the shock at each instant and find the altered heating rates and
the consequent changes in NSE self-consistently. In this letter we have
estimated the heating rates
for full flavour conversion between the neutrino-sphere and the shock.

We consider matter consisting of nucleons as well as nuclei and
incorporate charge current cross-sections on nuclei from \cite{haxton},
where these are listed upto a temperature of 6 MeV. If
flavour conversion takes place, the $\nu_e$ temperature can be higher.
A simple-minded extrapolation yields a much higher
cross-section and consequently a higher heating.
We take a conservative standpoint and use the cross-section given for
6 MeV as the value at higher energies. We also incorporate the
neutral current inelastic scattering on nuclei. As pointed out in
\cite{haxton} and later shown explicitly in \cite{rk92}, such
reactions can increase the shock heating rates substantially. However,
these processes being flavour independent, the heating due to them is
unaffected by neutrino oscillations.
The energy absorbed by matter behind the shock front/gm/sec is given by,
$\dot{E}~ =~ \dot{E}_{1}~ -~ \dot{E}_{2}~ + ~\dot{E}_{scatt}$,
$\dot{E_{1}}$ is the heating rate due to charged current
absorption of $\nu, \bar{\nu}$ on free nucleons and nuclei.
$\dot{E_{2}}$ is the rate of energy loss
due to neutrino radiation at matter temperature $T_{m}$ and
$\dot{E}_{scatt}$ is due to $\nu$-e scattering
and is ${T_m}/T_{\nu}$ times the difference of the first two terms
\cite{bw85} plus the contribution due to inelastic neutral current
scattering on nuclei \cite{haxton}.
With complete flavour transformation these terms are different from
what they were in the absence of oscillation \cite{bw85,haxton},
$T_{\nu_e}$ in these terms being replaced by $T_{\nu_{\tau}}$.
The resulting increase in $\dot{E}$ is quite significant.
The energy absorbed first goes into dissociating the iron nuclei into
$\alpha$-particles via $^{56}Fe \rightleftharpoons 13\alpha + 4n$. The
$\alpha$-particles then break up into protons and neutrons through
$\alpha \rightleftharpoons 2p + 2n$. The Q-values per nucleon for these
are 2.2 MeV and 7.075 MeV respectively. The exact nuclear composition at
each instant of time is governed by the Saha equations \cite{rk87}.
The change in entropy $\Delta S$ due to the
absorption of energy and changes in nuclear composition in
time $\Delta t$ is given, in the limit of zero mass accretion, by
\begin{equation}
T_{m}{\Delta}S = \dot{E}{\Delta}t - 2.2 {\Delta}X_{Fe} -
7.075{\Delta}X_{\alpha}
\label{tds}
\end{equation}
In this $X_{i}$ denotes the relative fraction of
species $i$, ${\Delta}X_{i}$
is the change in time ${\Delta}t$. For non-zero mass
accretion there will be additional terms in (\ref{tds}) due to
the kinetic energy of the accreting matter plus the terms due to
compositional changes of the matter as it moves across the shock front
The corresponding change in temperature ${\Delta}T$ is obtained as in
\cite{rk92}. The internal energy of the matter per nucleon
${\rm \epsilon}_{int}$ has contributions coming from baryons,
electrons and photons.
The success or faliure of this phase of heating depends on the
comparative values of ${\rm \epsilon}_{int}$ and ${\rm \epsilon}_{cr} =
GM_{c}/R_{c}$, where $R_{c}$ denotes the bifurcation radius and $M_{c}$
is the mass included within $R_{c}$.

For illustrative purposes numerical calculations are performed for a
$25 M_{\odot}$ star \cite{wilson} with an initial iron core mass of
$1.37M_{\odot}$. The electron
neutrino luminosities ($L_{\nu_e}$) for which calculations are
performed are $4 \times 10^{52}$ and $2 \times 10^{52}$
ergs/sec with the corresponding temperatures of neutrino-spheres as
4.45 and 5.3 MeV. The luminosities of neutrinos and antineutrinos of all
flavours are assumed to be the same. The ${\nu_e}$-sphere radius,
$R_{\nu}$,
is 30 km. The  temperatures of the $\mu$ and $\tau$ neutrino-spheres are
taken as 10 MeV. The shock stalls initially at a radius of 480 kms. It
then recedes upto 189 kms and moves forward again \cite{rk92}. We consider
non-zero mass accretion with the rate  0.1 $M_\odot$/sec.

Demanding that a resonance is achieved between the neutrino-sphere and
the shock so that $R_{\nu} \leq r_{res} \leq R_{m}$, an idea regarding
the bounds on ${\Delta}m^2$ can be obtained
under the assumption of small mixing angles.
For complete flavour conversion one has to ensure that not only is a
resonance attained, but $P(\nu_f \rightarrow \nu_f)$ as given by
(\ref{pnuf}) is zero.
(99 - 100)\%  flavour conversion of all neutrinos in the energy range
1-100 MeV requires
$1.13 \times 10^{4} eV^2 \leq {\Delta}m^2 \leq 1.62 \times 10^{4} eV^2$,
$1.83 \times 10^{-7} \leq {{\sin^2}2\theta_V}/{\cos{2\theta_{V}}} \leq
7.85 \times 10^{-3}$. This is the most stringent bound on the
parameter values. However, heating is most
effective for neutrino energies 10-35 MeV. For such
neutrinos the allowed area in the parameter space is
$2.95 \times 10^{3} eV^2 \leq {\Delta}m^2 \leq 10^{5} eV^2$,
$1.45 \times 10^{-8} \leq {{\sin^2}2\theta_V}/{\cos{2\theta_{V}}} \leq
3.39 \times 10^{-2}$.
This is consistent with the cosmological limit of
$m_{\nu_{\tau}} \leq$ 100 eV \cite{cowsik}.
The nonadiabatic MSW solution to the solar neutrino problem,
which requires
$\Delta{m^2} \sim 6 \times 10^{-6} eV^{2}$ and $\sin^{2}2\theta_{V}
\sim 7 \times 10^{-3} $
\cite{lan} is due to $\nu_{e}-\nu_{\mu}$ oscillations in this
framework.

Figure 1a shows the time evolution of the fraction of
$\alpha$-particles, neutrons and protons with and without
oscillation for $L_{\nu_e}$ = $4 \times 10^{52}$ ergs/sec.
The starting point of our calculations
t=0.417 sec is the time at which the shock
starts retreating, where t=0 corresponds to core bounce.
By this time iron is completely dissociated for the $25 M_{\odot}$ star
considered. The energy delivered to the matter goes to dissociate the
$\alpha$-particles present. As is seen from figure 1a, in the presence of
neutrino flavour conversion, dissociation of $\alpha$'s takes place at
a faster rate because of the enhanced heating.
This further  raises the
heating rate due to an increased free nucleon fraction.
Figure 1b shows the total internal energy per nucleon as a function of
time for two representative luminosities
$L_{\nu_e}$ = $4 \times 10^{52}$ ergs/sec, $2 \times 10^{52}$ ergs/sec.
For a $25 M_{\odot}$ star, for the two chosen luminosities,  even
without oscillation ${\rm \epsilon}_{int}$ is greater than the critical
energy ${\rm \epsilon}_{cr}$.
For $L_{\nu_e}$ = $4 \times 10^{52}$ ergs/sec, neutrino oscillation
increases it by 61.33\% at the end of this heating phase.
In case of $L_{\nu_e}$ = $2 \times 10^{52}$ ergs/sec,
${\rm \epsilon}_{int}$ was just above the critical value but
complete flavour conversion raises it by 56.37\%.

The results of \cite{bw85} or \cite{mayle} indicate that this phase of
shock revival is followed by the creation of a quasi-vacuum, dominated
by radiation, which ends with an eventual expulsion of matter. Our initial
results indicate that resonant neutrino oscillations may indeed be
the important ingredient, hitherto missing, in the delayed neutrino
heating scenario to give a stronger shock in conformity with observation.
One needs to extend these calculations to later times
as well as incorporate the physics of neutrino oscillations in
hydrodynamic codes. We are also studying the effect of these
oscillations on a $18M_{\odot}$ star \cite{mayle} with neutrino
luminosities as given in \cite{burrows} for its importance in
connection to SN1987A.
Earlier calculations for a $18M_{\odot}$ star without neutrino
oscillations failed to strengthen the shock adequately \cite{rk92}.
Our preliminary calculations show that with matter induced neutrino
oscillation the internal energy becomes greater than the critical energy
in this case as well. In conclusion, we stress that resonant neutrino
oscillations open up exciting possibilities for type-II supernove through
the delayed neutrino heating mechanism and merit detailed examination.

\vskip 30pt
\parindent 0pt
The authors are indebted to Alak Ray for discussions and help.
S.G. is supported
by the Council of Scientific and Industrial Research, India while
A.R. has been supported in part by the Council of Scientific and
Industrial Research, India and the
Department of Science and Technology, India.

\newpage
\begin{center}
FIGURE CAPTION
\end{center}
Fig.1. Time evolution of (a). X$_{p}$, X$_{n}$, X$_{\alpha}$
for a $25M_{\odot}$ star with $R_{\nu}$ = 30 km.
The solid (dashed) curves are with (without) neutrino oscillations.
(b). The internal energy per nucleon ${\rm \epsilon}_{int}$
of a mass shell behind the shock front. Curves I (II) are for luminosities
$L_{\nu} =2 \times 10^{52}$ ($L_{\nu}$ = $4 \times 10^{52}$) ergs/sec.
Also shown is the critical energy ${\rm \epsilon}_{cr}$.
\end{document}